\documentstyle[prb,eqsecnum,aps,epsf]{revtex}
\begin{document}
\draft

\twocolumn[\hsize\textwidth\columnwidth\hsize\csname@twocolumnfalse\endcsname

\title{
Laughlin-Jastrow-correlated Wigner crystal in a strong magnetic field
}

\author{Hangmo Yi and H.A.\ Fertig}

\address{
Department of Physics and Astronomy and Center for Computational Sciences, University of Kentucky, Lexington, KY 40506
}

\date{\today}

\maketitle

\begin{abstract}
We propose a new ground state trial 
wavefunction for a two-dimensional Wigner crystal
in a strong perpendicular magnetic field.  
The wavefunction includes Laughlin-Jastrow correlations between 
electron pairs, and may be interpreted as a crystal state of
composite fermions or composite bosons.  Treating the 
power $m$ of the Laughlin-Jastrow factor as a variational 
parameter, we use quantum Monte Carlo simulations 
to compute the energy of these new states.  
We find that our wavefunctions have lower energy than 
existing crystalline wavefunctions in the lowest Landau 
level.  Our results are consistent with experimental
observations of the filling factor at which the transition between 
the fractional quantum Hall liquid and the Wigner crystal occurs
for electron systems.  
Exchange contributions to the wavefunctions are estimated quantitatively
and shown to be negligible for sufficiently small filling factors. 
\end{abstract}
\pacs{PACS numbers: 73.40.Hm, 73.20.Dx}
]

\section{Introduction}

It was first argued by Wigner in 1934 that a system of interacting but 
otherwise structureless electrons can have crystalline order 
in the limit of low density and low temperature \cite{wigner34}.  
The first experimental 
evidence of the Wigner crystal (WC) was found well over 40 years later 
in a two-dimensional (2D) system of 
electrons adsorbed on a helium surface \cite{grimes79}.  Nowadays, 
semiconductor heterojunction devices are considered a very promising 
environment for observing the WC.  The advantage of 
heterojunction structures comes from the fact 
that the 2D electron plane is spatially 
separated from the donor layer, so that the 
influence of these impurities can be substantially 
reduced compared to bulk semiconductor 
environments.  Furthermore, it is now well known that 
a strong magnetic field perpendicular to the 
2D plane can effectively localize electron 
wavefunctions while keeping the kinetic energy 
controlled \cite{yoshioka79,yoshioka83,fertig97}.  
Since this lessens the otherwise severe low-density 
condition, it is believed that the WC can be stabilized 
in a sufficiently strong magnetic field.  

On the other hand, the fractional quantum Hall (FQH) liquid 
is known to be the ground state in certain ranges of 
strong magnetic field \cite{tsui82,prange90}.  In this 
strongly correlated liquid state, the Hall 
resistivity $\rho_{xy}$ is quantized at discrete values 
and the diagonal resistivity $\rho_{xx}$ vanishes 
at zero temperature.  In contrast, $\rho_{xx}$ presumably 
diverges at $T=0$ in the pinned WC.  The FQH effect 
for $\nu=1/m$ ($m$ odd) is now fairly well understood in
terms of the 
Laughlin wavefunction \cite{laughlin83}.  

In several recent experiments with high mobility samples, a sharp 
phase transition from the FQH state to an insulating state 
was observed as the magnetic field was increased both in 
electron \cite{jiang90,goldman90} and hole \cite{santos92} systems.  
Some properties of the insulating state such as the activation 
gap in charge transport closely resembles those of  
pinned charge density waves, supporting the interpretation 
of this insulating state as a WC.  
Theoretical calculations of both the FQH liquid energy \cite{price94} 
and the WC energy \cite{zhu93,lam84} are 
also in good agreement with the experiments as to the critical 
value of the magnetic field at which the transition occurs 
for a given electron density.  

However, {\em not} all experimental findings of the 
insulating state are consistent with the conventionally accepted 
theoretical understandings of the WC.  (I) First, there is 
a discrepancy in the energy of charged excitations.  Particularly, 
transport experiments \cite{jiang90} reveal 
that the activation gap is an order 
of magnitude smaller than the theoretically estimated energy to 
create a point defect in the WC \cite{fisher79,cockayne91,price91}.  
(II) Moreover, even deep into the insulating phase, anomalous 
behavior is observed when the filling factor $\nu$ is an 
inverse odd integer, which may be related 
to the FQH effect.  Specifically, transport experiments \cite{goldman88} 
exhibit a dip in the diagonal resistivity $\rho_{xx}$ of the 
insulating state near $\nu=1/7$.  Also, photoluminescence 
experiments \cite{buhmann91,clark91,kukushkin92,goldys92}
exhibit structure near odd denominator filling
factors down to
1/11, which looks very similar 
to structure seen at higher fillings where the FQH effect occurs.
(III) Finally, experiments \cite{goldman88,goldman93,sajoto93} show 
that the Hall resistivity $\rho_{xy}$ in the insulating phase saturates 
at its classical value $B/nec$, just as in the FQH liquid 
phase.  This behavior {\em cannot} be understood in terms of a 
model of thermally activated point defects that are essentially 
non-interacting \cite{zheng94b}.  Interestingly, (II) and (III) 
suggest that some characteristics of the FQH effect are shared 
by the insulating state.  

The unusual behavior of $\rho_{xy}$ has led to 
speculation that the insulating phase is not a WC at all, 
but rather a disorder-dominated state called the ``Hall 
insulator'' \cite{kivelson92,zhang92}.  However, it has been shown that 
interstitial defects in a WC can also lead to Hall insulating 
behavior if one introduces Laughlin-Jastrow correlations 
between the interstitials and the lattice 
electrons \cite{zheng94b,zheng94}.  The correlation 
was found to lower the energy to create such defects.  
However, more careful studies \cite{unpub} of the above interstitial 
state using Monte Carlo simulations 
suggest that in order to obtain such a small excitation energy 
as found in experiment, one must introduce Laughlin-Jastrow 
correlations into the ground state as well.  
In this paper we explore the energetics of ground state
wavefunctions of this form.

In what follows we will introduce trial wavefunctions that
take the form of a Laughlin-Jastrow factor multiplying a properly
(anti-)symmetrized product of single particle states.
The wavefunction introduced here thus corresponds to 
composite fermion or boson states \cite{jain,fisher}.
The energies of these states are computed using
quantum Monte Carlo simulations, and it will be
demonstrated that such states are generically lower
in energy than other lowest Landau level WC states
in the literature.  Our computational method in its
simplest form ignores exchange corrections; i.e., 
the state multiplying the Laughlin-Jastrow factor
is approximated as a simple product of single particle
states.  An in-principle exact computational scheme
in which permutations of the single particle states
are sampled shows that this is an excellent approximation,
provided the filling factor is not too large.

This paper is organized as follows: The trial many body 
wavefunction of the Laughlin-Jastrow-correlated WC 
is introduced and some of its properties are discussed 
in Sec.~\ref{sec:wavefunction}.  In 
Sec.~\ref{sec:energy}, the ground state energy 
is computed using a Monte Carlo simulation.  Various 
aspects of the results are also discussed.  
Sec.~\ref{sec:exchange} is devoted to discussions on the 
effect of the exchange energy and validity of our 
approximation.  Finally, we 
summarize the findings in Sec.~\ref{sec:summary}.  
Some technical details of the energy calculation can be 
found in the Appendix.

\section{Trial Wavefunction}
\label{sec:wavefunction}

The Hamiltonian of 2D electrons moving in a magnetic field ${\bf B}$ 
and interacting with the Coulomb energy is written
\begin{equation}
H = \sum_i {1\over 2m} \left| {\bf p}_i-{e\over c}{\bf A}({\bf r}_i) \right|^2 + {1\over 2} \sum_{i\neq j} {e^2\over|{\bf r}_i-{\bf r}_j|},
\end{equation}
where ${\bf r}_i$ and ${\bf p}_i$ are the 2D position 
and momentum of the $i$-th electron, and ${\bf A}$ is 
the vector potential from which the magnetic field is 
given by ${\bf B} = \nabla \times {\bf A}$.  
We will ignore the spin degree of freedom assuming that it is 
completely polarized by the strong magnetic field.  

Since our trial wavefunction is closely related to the Hartree-Fock 
wavefunction of the WC suggested in Ref.~\ref{ref:maki83}, 
\nocite{maki83} it is 
worthwhile to summarize the aspects of the Hartree-Fock 
wavefunction here.  Explicitly, it can be written
\begin{equation}
\Psi_{HF}(\{{\bf r}_i\}) = {\cal A} \prod_i \phi^{HF}_{{\bf R}_i}({\bf r}),
\end{equation}
where $\phi^{HF}_{{\bf R}_i}$ are single electron wavefunctions 
and ${\cal A}$ antisymmetrizes the total many body wavefunction.  
To a first approximation in the large magnetic field, all electrons 
will lie completely in the lowest Landau level.  The un-normalized 
single particle wavefunction is thus given by
\begin{equation}
\phi^{HF}_{{\bf R}_i}({\bf r}) = e^{-{|{\bf r} - {\bf R}_i|^2\over 4\ell^2} - i{{\bf r}\times{\bf R}_i\cdot\hat{z}\over 2\ell^2} }.
\label{eq:singleHF}
\end{equation}
This wave function describes an electron localized 
at ${\bf R}_i$ within a 2D Gaussian wavepacket.  
The magnetic length $\ell=\sqrt{\hbar c/eB}$ determines 
the size of the wavepacket.  The phase factor in 
Eq.~(\ref{eq:singleHF}) ensures that $\phi^{HF}_{{\bf R}_i}$ 
is a product of $e^{-|{\bf r}|/4\ell^2}$ and an analytic 
function of $z\equiv x+iy$, making it entirely lie 
in the lowest Landau level.  In the large $B$ 
limit, $\ell\rightarrow 0$, and the electrons become 
highly localized, behaving almost like classical point charges.  
The kinetic energy is given by the lowest Landau level 
energy $\hbar\omega_c/2\equiv\hbar eB/2mc$, and is 
the same regardless of the Gaussian center ${\bf R}_i$.  
This allows one to treat ${\bf R}_i$ 
as variational parameters in minimizing the total energy 
with respect to the Coulomb interaction.  For a classical system, 
a triangular lattice is well known to have the lowest Coulomb 
energy for a given density \cite{bonsall77}.  
Therefore, in the limit $B\rightarrow\infty$, 
the ground state is expected to be represented by the above 
wavefunction, with ${\bf R}_i$ forming a triangular lattice.  

In a finite magnetic field, quantum fluctuations around the 
lattice sites become important and the above classical analogy 
is only approximate.  Consequently, $\Psi_{HF}$ is not 
guaranteed to give the lowest energy at finite ${\bf B}$.  
However, if ${\bf B}$ is large enough, $\Psi_{HF}$ is still 
very close to the true ground state.  For this reason, 
it has been used even at finite ${\bf B}$ by many authors, 
producing very good results.  In this paper, however, we 
will improve upon it by introducing a correlation. 
Previous studies \cite{lam84} have introduced correlation
factors that are exact for a harmonic Hamiltonian;
however, such wavefunctions allow fluctuations in which
particles may occasionally closely approach one another.
A correlation factor which by now is well-known to
suppress such fluctuations is the Laughlin-Jastrow factor:
$\prod_{i<j} (z_i-z_j)^m$,
where $z_i=x_i+iy_i$ is the complex notation of the electron 
coordinates.  
Because of the extra phase
accumulated when one particle encircles a second,
wavefunctions of this form may be understood
as being comprised of particles that have $m$
magnetic flux quanta attached to them.  The idea
of constructing wavefunctions of this general form was first
suggested in the context of the FQH effect in groundbreaking
work by Jain \cite{jain}; the combination of electrons and
an even number of flux quanta to form these wavefunctions
have since become known as composite fermions.  For odd
values of $m$, the wavefunction multiplying the Jastrow
factor must be symmetric under interchange of two particles,
so that such states correspond to composite boson states \cite{fisher}.
The wavefunctions we study in this work may thus be interpreted
as crystals of composite fermions or bosons.

We therefore propose the following trial wavefunction:
\begin{equation}
\Psi(\{{\bf r}_i\}) = {\cal A} \prod_{i\neq j} (z_i-z_j)^m \prod_i \phi_{{\bf R}_i}({\bf r}_i)
\label{eq:mbwf}
\end{equation}
Again, $\phi_{{\bf R}_i}$ is a single particle 
wavefunction which is localized at ${\bf R}_i$ and lies in 
the lowest Landau level.  Since we 
will consider finite size systems, only those lattice sites within 
a disk of radius $R_D$ will be included in the set 
of $\{{\bf R}_i\}$.  In order to make the full 
wavefunction antisymmetric, we choose to use 
either a Slater determinant (even $m$) or 
a symmetric sum of all possible permutations (odd $m$) 
for the single particle wavefunction part.  

We cannot, however, simply use $\phi^{HF}_{{\bf R}_i}$ 
in Eq.~(\ref{eq:singleHF}) as our single particle wavefunctions, 
due to the following reason.  
Disregarding the antisymmetrization for the moment, the 
probability density is given by
\begin{eqnarray}
|\Psi|^2 & \sim & \prod_{i\neq j} |{\bf r}_i-{\bf r}_j|^{2m} \prod_i |\phi_{{\bf R}_i}({\bf r}_i)|^2 \\
& = & e^{2m \sum_{i\neq j} \log |{\bf r}_i-{\bf r}_j|} e^{2\sum_i \log |\phi_{{\bf R}_i}({\bf r}_i)|}. \label{eq:plasma}
\end{eqnarray}
As in Laughlin's ``plasma analogy'' \cite{laughlin83}, $|\Psi|^2$ 
may be thought of as the Boltzmann distribution function for 
a ``dual'' classical system whose effective energy is 
given by the exponents in the above equation, 
up to an arbitrary effective temperature.  
The first exponent in Eq.~(\ref{eq:plasma}) is identical 
the 2D (logarithmic) Coulomb energy with coupling 
constant $m$.  Each term in the second exponent describes 
an attractive effective potential centered at a lattice 
site ${\bf R}_i$.  Obviously, if the effective Coulomb 
interaction were absent, the minimum of the effective energy 
would be achieved when ${\bf r}_i={\bf R}_i$ for all $i$.  
However, due to the effective interaction, the static 
solution of ${\bf r}_i$ will be moved away from ${\bf R}_i$, 
unless ${\bf R}_i$ is the center of the disk.  In general, 
the electrons will be pushed radially away from 
the center of the disk.  Consequently, the whole system 
will spread out and the resulting electron density 
will be smaller than that of the intended lattice.  

\begin{figure}[tb]
\epsfxsize=2.6in
\centerline{ \epsffile{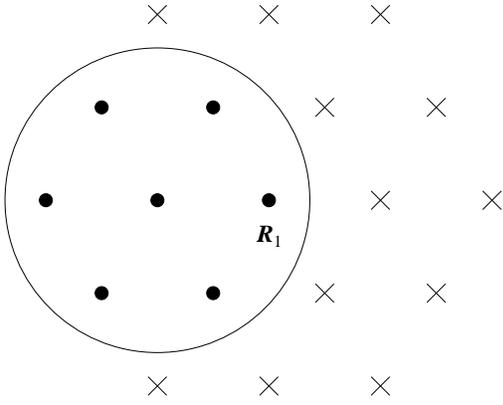} }
\caption{
An example of the underlying triangular lattice for a finite size 
WC.  The large circle denotes the boundary of the physical disk and 
the filled dots the lattice sites, ${\bf R}_i$, 
within the disk.  The crosses 
denote the positions of the ghosts, $\eta^{(1)}_j$, which 
balance the effective force at ${\bf R}_1$.  
}
\label{fig:ghosts}
\end{figure}

In order to prevent this unwanted expansion of the system, 
for each $\phi_{{\bf R}_i}$, we will introduce extra zeros 
(``ghost effective charges'') outside the physical disk.  
The ghosts are introduced in such a way that if the 
real electrons were fixed at their lattice sites, 
the total of both real and ghost effective charges 
are symmetrically distributed about any given lattice 
site ${\bf R}_i$.  In other words, the ghosts cause each lattice 
site to look like it is at the center of the system by 
``balancing'' out the effective repulsive force 
of the surrounding electrons.  As a consequence, 
each electron will remain centered near its own lattice 
site.  Obviously, the specific positions of 
the ghosts depend on the lattice site ${\bf R}_i$. 
An example of the way ghosts are 
placed is shown in Fig.~\ref{fig:ghosts}.  Although 
outside the physical disk, the ghosts themselves occupy 
lattice sites.  Including the ghosts, the single 
particle wavefunction is finally given by
\begin{equation}
\phi_{{\bf R}_i}({\bf r}) = e^{-{|{\bf r} - {\bf R}_i|^2\over 4\ell^2} - i{{\bf r}\times{\bf R}_i\cdot\hat{z}\over 2\ell^2} } \prod_j (z-\eta^{(i)}_j)^m
\label{eq:single}
\end{equation}
where $\eta^{(i)}_j$ are the complex coordinates of the 
ghosts that balance out the effective force at ${\bf R}_i$.  

An interesting property of these wavefunctions is that
the ghosts may be thought of as 
``renormalizing'' the positions of lattice sites.
To see this, one can rewrite 
Eq.~(\ref{eq:single}) as
\begin{eqnarray}
\phi_{{\bf R}_i}({\bf r}) & = & e^{-{|{\bf r} - {\bf R}_i|^2\over 4\ell^2} - i{{\bf r}\times{\bf R}_i\cdot\hat{z}\over 2\ell^2} } \nonumber \\
 & & \times \exp \bigg[ m \sum_j \log |z - \eta^{(i)}_j| \nonumber \\
 & & \quad \qquad + im\sum_j \mbox{arg} (z - \eta^{(i)}_j) \bigg].
\end{eqnarray}
Once again in the plasma analogy, the logarithms in the 
exponent describe a 2D Coulomb potential caused by 
effective point charges with charge $m$.  Now let us 
approximate the point charges $\{\eta_j\}$ 
by a uniform charge 
distribution whose density is the same as the average density 
of the original point charges.  
To do this, we write the sums in the argument of the
exponential in the form
\begin{eqnarray}
& & m\sum_j \{ \log|z-\eta_j| + i \arg(z-\eta_j) \} \nonumber \\
& & \quad \equiv m \int d^2\eta \{ \log|z-\eta| + i \arg(z-\eta) \} \rho_G(\eta)
\end{eqnarray}
where $\rho_G(\eta)$ is the density of ghost particles.
We then approximate
\begin{equation}
\rho_G(\eta) \approx \left\{
  \begin{array}{ll}
  \overline{\rho}\ \ & \text{if $|\eta|>R_D$ and $|\vec{\eta} - \vec{R}_i| < R_G$}, \\
  0 & \text{otherwise,} \\
  \end{array} \right.
\end{equation}
where $\overline{\rho}$ is the average electron density, $R_D$ is
the radius of the physical disk of the finite size system, and
$R_G$ is the radius of a ``ghost disk'', which must satisfy
$R_G>2R_D$.  Provided $r$ is well away from the physical
disk edge, this approximation should be quite good, and
we expect corrections to scale as $(r/R_D)^2$.  Since
the real part of the integral corresponds to the potential
of a uniform charge density $m\overline{\rho}$ in a disk
of radius $R_G$, with a circular hole of radius $R_D$,
the real part of the integral may be computed using
Gauss' law for two-dimensional electrostatics.  The imaginary
part of the integral may be computed analytically as well
for $r \ll R_D$, yielding the approximated wavefunction
\begin{eqnarray}
\phi'_{{\bf R}_i}({\bf r}) & = & e^{-{|{\bf r} - {\bf R}_i|^2\over 4\ell^2} - i{{\bf r}\times{\bf R}_i\cdot\hat{z}\over 2\ell^2} } \nonumber \\
 & & \times \exp \bigg[ {\pi m\overline{\rho} \over 2}\left(|{\bf r} - {\bf R}_i|^2-|{\bf r}|^2\right) \nonumber \\
 & & \quad \qquad - i\pi m\overline{\rho} {\bf r}\times{\bf R}_i\cdot\hat{z} \bigg] \\
& = & e^{-{|{\bf r} - (1-m\nu) {\bf R}_i|^2\over 4\ell^2} - i{{\bf r}\times(1-m\nu){\bf R}_i\cdot\hat{z}\over 2\ell^2}} e^{-{m\nu(1-m\nu)|{\bf R}_i|^2\over 4\ell^2}}. \nonumber \\
\end{eqnarray}
Note that because the amplitude and phase of $\phi'_{{\bf R}_i}$
have been treated on an equal footing, this
wavefunction lies in the lowest Landau level.  Ignoring 
the unimportant constant, $\phi'_{{\bf R}_i}$ 
describes an electron in the lowest Landau level, centered 
at a renormalized lattice site $(1-m\nu) {\bf R}_i$.  
Thus the ``bare'' 
lattice described by filling $\phi'_{{\bf R}_i}$ states
will be smaller than the real lattice by a factor of
$1-m\nu$.  The physical lattice however is
spread back to its original size due to the 
Laughlin-Jastrow correlation.  Therefore, the above 
``renormalization'' of the lattice  
compensates for the previously mentioned lattice expansion 
due to the Laughlin-Jastrow factor.

Before we describe the energy calculation for our wavefunctions, let 
us briefly discuss about the effect of the Laughlin-Jastrow 
correlation to the characteristics of the WC, particularly 
in connection with the excitation energy.  In a recent 
experiment \cite{jiang90}, Jiang et al., have measured the 
temperature dependence of the diagonal resistance in the 
reentrant insulating phase slightly above $\nu=1/5$.  
According to their data, the activation gap for 
charge transport is given by $E_g\sim 0.63$K.  Surprisingly, 
this energy is much smaller than would be theoretically expected.  
For example, using states of point particles whose positions are 
chosen to optimize the energy, the energy to create a point 
defect such as an interstitial or a vacancy has been estimated 
by many authors \cite{fisher79,cockayne91,price91}, 
but all the results are an order of 
magnitude greater than the above value of $E_g$.  
A more recent study of point defects, however, shows that 
the energy can be lowered if the Laughlin-Jastrow correlation 
is introduced between the interstitials and the 
lattice electrons \cite{zheng94}.  
Our initial studies of interstitial wavefunctions using quantum
Monte Carlo techniques such as those presented here suggest
that to reach the very low activation energies seen in
experiment, one needs to include Laughlin-Jastrow correlations
among the ground state electrons as well \cite{unpub,unpub2}.  
A discussion of such wavefunctions is deferred to a future 
publication.

\section{Coulomb Energy: Monte Carlo Simulation}
\label{sec:energy}

Since our wavefunction lies completely in the lowest Landau level, 
we only need to minimize the Coulomb interaction term in the 
Hamiltonian.  The expectation value of the Coulomb energy per 
electron is written
\begin{eqnarray}
{E_c\over N} & = & {1\over 2N} \sum_{i\neq j} \left< e^2\over|{\bf r}_i-{\bf r}_j| \right> \\
& = & {e^2\over 2N} \int d{\bf r}d{\bf r}' {\left< \sum_{i\neq j} \delta({\bf r}-{\bf r}_i)\delta({\bf r}'-{\bf r}_j) \right> \over|{\bf r}-{\bf r}'|}, \\
& = & {e^2\over 2} \int_C d{\bf r} \int d{\bf r}' {\left< \sum_{i\neq j} \delta({\bf r}-{\bf r}_i)\delta({\bf r}'-{\bf r}_j) \right> \over|{\bf r}-{\bf r}'|},
\label{eq:ec}
\end{eqnarray}
where $\langle\cdots\rangle$ means the expectation value with respect 
to the wavefunction $\Psi$ in Eq.~(\ref{eq:mbwf}).  In the last line, 
we have dropped $1/N$ and restricted the first 
integral within a single primitive cell at the center of the 
disk (denoted by $C$), using the lattice symmetry.  
Since the size of the simulated system is inevitably finite, 
in order to obtain the thermodynamic limit, we need to either 
extrapolate finite size results, or use the Ewald sum 
method \cite{bonsall77,ewald}.  We have used the second method 
in this paper.  Details of the calculation 
are given in Appendix A, but it 
must be noted here that we have introduced a couple of approximations 
in calculating the Coulomb energy: (I) We have ignored the 
exchange energy, 
which in practice means that the antisymmetrization
in Eq.~(\ref{eq:mbwf}) is dropped.
We have tested this approximation and find that
it is quite good unless $\nu$ 
is too close to $1/m$.  The effect of exchange energy will be 
discussed in the next section in more detail.  
(II) Since $|\Psi|^2$ when unsymmetrized corresponds to
a finite temperature classical Boltzmann weight, we assume
there exists a length scale $\xi_c$ above which fluctuations
in the electron positions are uncorrelated.  We thus use the
Monte Carlo method to compute the Coulomb interaction
between the charge density in the central unit cell
and the charge out to some distance $R_S$ which we
presume to be larger than $\xi_c$.  This run is also
used to compute the charge density in the central
primitive unit cell.  In order to minimize boundary 
effects, we choose $R_D$, the radius of the disk containing all 
simulated dynamical electrons, to be greater than $R_S$, 
so that electrons close to $R_S$ do not experience an 
environment significantly different than those in the 
bulk.  Those electrons between radius $R_S$ and $R_D$, 
which are dynamically simulated but not used to 
compute the energy, provide an ``effective medium''.  
This approach has also been employed in Monte Carlo 
studies of the FQH effect \cite{morf86}.  

The interaction of the charge density
in the central unit cell with charge at distances greater
than $R_S$ is computed by treating the distant charge as
static and equal to periodic copies of the 
numerically computed charge density in the central 
primitive cell.  This is essentially
a Hartree approximation.  Since this charge density is treated
as static, one may compute the interaction for an infinitely
large system using the Ewald sum technique.  Our method is checked
by increasing $R_S$ until the energy is unchanged within the
errorbars of our Monte Carlo calculations.  Our simulations 
show that for the wavefunction parameters we have studied, $\xi_c$ 
is always less than $4a$, where $a$ is the lattice constant.  
This is also confirmed by numerical calculations of individual 
pair energies, for which the result from the simulation 
is essentially the same as the Hartree energy if 
the pair is separated farther than $4a$.  More details of 
this procedure are discussed in Appendix A.  

\begin{figure}[tb]
\epsfxsize=3.2in
\centerline{ \epsffile{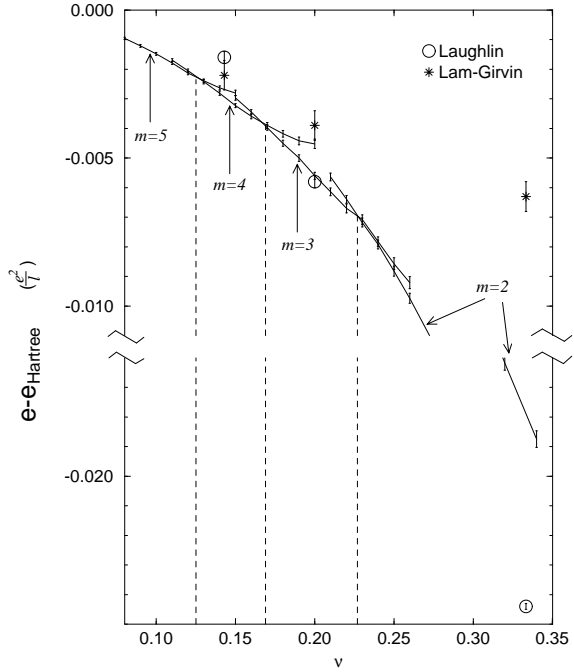} }
\caption{
The Coulomb energy per electron of the Laughlin-Jastrow-correlated 
Wigner crystal as a function of $\nu$ for various values of $m$.  
The energy is shown relative to that of the uncorrelated 
Hartree wavefunction ($m=0$).  
The same quantity is presented for the Laughlin state 
(Ref.~\protect\ref{ref:lam84}) 
and the Lam-Girvin wavefunction (Ref.~\protect\ref{ref:levesque84}) 
at $\nu=1/3$, $1/5$, and $1/7$.  
The dashed vertical lines represent the values of $\nu$ where 
the transition between different $m$ states occur.  
}
\label{fig:energy}
\end{figure}

We have developed a Monte Carlo simulation program that computes 
the Coulomb energy per electron, $E_c/N$, using the 
standard Metropolis algorithm \cite{metropolis}.  As a 
critical test of our extrapolation technique, we have used our 
method to compute the energy of the $m=0$ state, which is 
identical to the one used in Ref.~\ref{ref:maki83}.  
Its energy can be calculated analytically and our results 
agree with analytic solutions well within the 
statistical errorbar of about 0.05\%.  
The results for more interesting values of $m$ are 
plotted in Fig.~\ref{fig:energy}\nocite{levesque84}.  
Treating $m$ as a variational parameter, one can find the 
value of $m$ which gives the lowest energy at a given $\nu$.  
The graph clearly shows that at $\nu=1/3$ and $1/5$, the 
Laughlin state has a lower Coulomb energy than any of 
our wavefunctions.  At $\nu=1/7$, however, the Laughlin 
state has a higher Coulomb energy than our lowest result.  This is 
consistent with experiment in that the ``true'' FQH effect --- 
e.g., vanishing diagonal resistivity $\rho_{xx}$ at 
zero temperature --- has never been observed at any inverse odd 
filling factors below $\nu=1/5$.  Furthermore, at $\nu=1/5$, 
the energy of 
our wavefunction is higher than, but very close to that of 
the Laughlin state, which agrees well with the observation of a
reentrant insulating phase \cite{jiang90} 
slightly above $\nu=1/5$.  This reentrant phase is believed to 
occur because the pure Laughlin wavefunction is the ground state 
only when $\nu$ is {\em precisely} an 
inverse odd integer.  Away from the precise filling factors, 
quasiparticles and quasiholes are present in the ground state, 
increasing the energy.  Therefore the FQH states have cusps in 
energy at every inverse odd filling factor, allowing the 
WC state to have lower energy in a small but finite range of $\nu$ 
right {\em above} $1/5$.  

Now, let us compare our results with other WC trial wavefunctions, 
particularly the Lam-Girvin form \cite{lam84}.  The Lam-Girvin 
wavefunction also predicts that the phase transition from the WC 
to the FQH effect occurs between $\nu=1/5$ and $1/7$.  As shown 
in Fig.~\ref{fig:energy}, however, our wavefunctions are 
lower in energy than the Lam-Girvin counterpart at all values 
of $\nu$ where data is available.  In other words, our wavefunctions 
are closer to the true ground state.   
We believe this difference 
arises because the harmonic approximation neglects rare, but 
nonetheless important contributions from 
anharmonic fluctuations in which two or more electrons come 
close together.  In contrast, the Laughlin-Jastrow correlation 
very effectively suppresses density fluctuations at all 
displacements of electrons from the lattice sites.  This 
may be understood using the plasma analogy for the Laughlin 
states \cite{laughlin83}, i.e., the Laughlin-Jastrow correlation 
is equivalent to the Boltzmann distribution of a 2D 
one component plasma (OCP) in 
which charge density fluctuations are suppressed.  

An important difference between the weighting associated with
our wavefunction and the Boltzmann weight of the OCP is that
the electrons are centered at different lattice sites in our 
wave function, while they are centered at one single 
point for the OCP.  One of the most significant 
consequences of this is the following.  Let us define $m_0(\nu)$ as 
the value of $m$ for the lowest energy variational state 
at $\nu$.  Surprisingly, near an inverse odd integer filling 
factor, we find $\nu \sim 1/(2m_0-1)$ 
rather than $\nu \sim 1/m_0$ as in the Laughlin states.  
For example, at $\nu=1/7$, our wavefunction has the lowest 
energy if $m=4$, rather than $m=7$.  Now let us continue to 
focus on the $m=4$ state increasing $\nu$ above $1/7$.  
It continues to be the lowest energy state until $\nu$ 
reaches about $\sim 0.165$, where the $m=3$ state becomes lower in 
energy.  This implies a first order transition 
between the two different $m$ states.  
This phase transition may in principle be detectable in 
photoluminescence experiments, although 
this is presumably difficult because the energies of 
neighboring $m$ states are so close together.  

\begin{figure}[tb]
\epsfxsize=3.2in
\centerline{ \epsffile{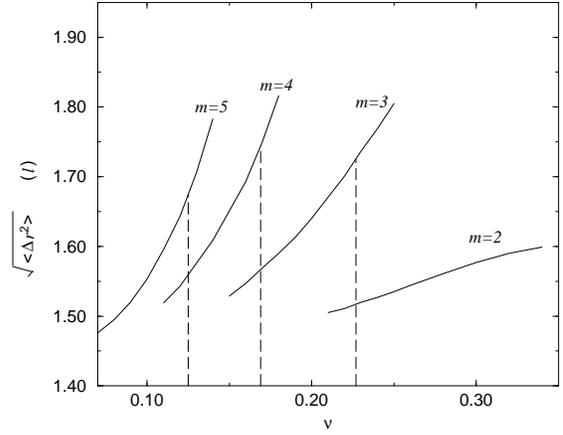} }
\caption{
Root-mean-square of the electron fluctuation from the 
lattice sites as a function of the filling factor $\nu$.  
The dashed vertical lines represent the values of $\nu$ where 
the transition between different $m$ states occur.  
}
\label{fig:rmsr}
\end{figure}

\begin{figure}[tb]
\epsfxsize=2.6in
\centerline{ \epsffile{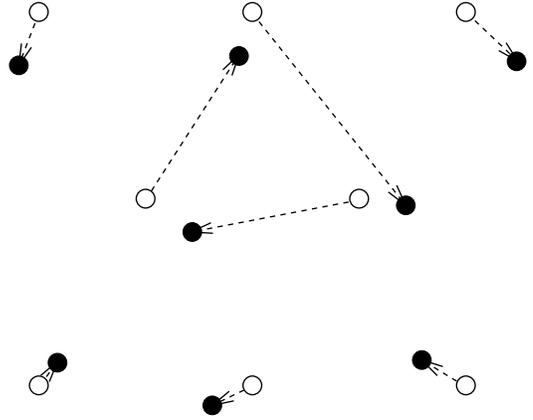} }
\caption{
A typical configuration of the three particle exchange ring 
which can be obtained from a ``snap shot'' of a Monte 
Carlo simulation.  Electron positions are denoted by 
filled circles and lattice sites by empty circles.  
The arrows indicate which electron is originated from 
which Gaussian center ${\bf R}_1$.  
}
\label{fig:ring}
\end{figure}

Fig.~\ref{fig:rmsr} shows $\Delta r_{\text{rms}}$, 
the root-mean-square 
value of the fluctuation of electrons from their lattice sites.  
Note that $\Delta r_{\text{rms}}$ 
increases rapidly as $\nu$ approaches and passes 
beyond the transition to the $m-1$ state.  This indicates 
that the single electron probability density becomes less and less 
localized as $\nu$ increases.  However, according to the 
above mentioned plasma analogy, the Laughlin-Jastrow 
factor still tries to force the local density to remain uniform.  
Therefore, rather than wandering around randomly, electrons tend 
to {\em switch} positions and form ``exchange 
rings'' (Fig.~\ref{fig:ring}).  
This means that ring exchange energy becomes more and more 
important.  This is most easily seen from ``snap shots'' of 
the electron configuration during a Monte Carlo simulation run. 
We note that such ring exchanges are commonly
observed in simulations of melting of the classical
OCP \cite{chouquard83}.  In path integral descriptions of the FQH 
effect \cite{ringexchange},
coherence among ring exchanges play a crucial role in
explaining the instability of the WC with respect to
a liquid state at $\nu=1/m$ for small enough $m$.
We believe that quantum
coherence in ring exchanges may lead to structure
in the energy of the WC as a function
of filling factor even in the insulating state, 
which ultimately could explain the transport and
photoluminescence anomalies discussed in the 
introduction.
However, a correct description of this requires that
exchange be properly included; we therefore defer 
a detailed discussion of this to a future publication \cite{unpub}.  
As $\nu$ approaches and increases past the critical 
filling factor, the exchange rings are observed 
increasingly often in Monte Carlo snap shots.  
As $\nu$ increases further and $\Delta r_{\text{rms}}$ 
grows to the same order of magnitude as the lattice constant, 
the WC will eventually become 
unstable, giving way to a liquid-like state.  This is analogous 
to the melting transition of a conventional solid.  In this limit, 
however, the exchange energy is clearly no longer negligible and our 
Monte Carlo analysis ceases to be valid.  Then, an important 
question arises: when may exchange be ignored?
We will address this question in the next section.  

Now let us focus on the transitions between different $m$ 
states.  First, Fig.~\ref{fig:rmsr} shows characteristics 
of $\Delta r_{\text{rms}}$ that is common to all $m$.  
In general, $\Delta r_{\text{rms}}$ is an increasing 
function of $\nu$, and as $\nu$ reaches some point, the system 
undergoes a phase transition to a lower $m$ state.  Interestingly, 
the values of $\Delta r_{\text{rms}}$ at the 
critical $\nu$ is approximately given by the same 
value $\sim 1.7\ell$ regardless of $m$.  We believe 
this implies that delocalization plays a crucial 
role determining where the transitions occur.  Moreover, 
Fig.~\ref{fig:energy} shows that $E_c/N$ starts to curve 
up as $\nu$ passes beyond the transition value, which is 
common for all $m$.  We believe the delocalization and the 
formation of exchange rings are the main reasons for this 
change of curvature in the Coulomb energy.  However, it is 
not yet clear why it occurs well below $\nu=1/m$.

\section{Exchange Effect: Permutation Monte Carlo Simulation}
\label{sec:exchange}

In the previous section, we have seen that the 
delocalization, which is represented 
by $\Delta r_{\text{rms}}$, increases as $\nu$ 
approaches $1/m$ from below.  Then the exchange energy is 
expected to become more and more important, as the overlap of 
wavefunctions at different lattice sites increase.  
Indeed, we have more or less directly observed the formation 
of exchange rings in the snap shots from the Monte Carlo simulations 
for relatively large values of $\nu$.  
Thus, it is clear that as $\nu$ grows, one must start to 
include the exchange energy in the calculation in order to obtain 
quantitatively reliable results.  

It is very difficult to estimate the exchange energy 
analytically for our wavefunction mainly due to the strong 
correlations.  However, when $m$ is even, the single particle 
wavefunction part in Eq.~(\ref{eq:mbwf}) is a Slater determinant, 
and we can take the exchange energy into full account 
by using other Monte Carlo methods such as in 
Ref.~\ref{ref:ceperley77}\nocite{ceperley77}.  Our 
tests with the $m=4$ state shows that the exchange energy is 
negligible when $\nu\lesssim 1/7$.  Although this method 
treats the exchange energy exactly, 
its application is strictly restricted to even values of $m$, 
and we need to resort to a different method for odd $m$.  

One way to estimate the relative importance of the exchange 
effect is as follows.  The many body wavefunction in 
Eq.~(\ref{eq:mbwf}) may be rewritten
\begin{equation}
\Psi = \prod_{i\neq j} (z_i-z_j)^m \sum_P \zeta^P \phi_{{\bf R}_i}({\bf r}_{P(i)}),
\end{equation}
where the summation is over all possible permutations $P$.  For 
an odd $m$, the statistical sign $\zeta^P$ is always $+1$, but 
for an even $m$, $\zeta^P$ 
is either $+1$ or $-1$ depending on whether $P$ is an even or odd 
permutation.  When 
we ignored the exchange effect in the previous section, what 
we did was to drop all permutations in the above summation, 
except the identity permutation $I$ such that $I(i)=i$.  
In other words, we have approximated the above wavefunction 
with
\begin{equation}
\Psi_{\text{direct}} = \prod_{i\neq j} (z_i-z_j)^m \phi_{{\bf R}_i}({\bf r}_i).
\end{equation}
Now, we want to define a quantity $\gamma$, which measures error 
caused by this approximation, or in other words, 
measures how important the exchange effect is.  First, 
the ``partition function'' maybe written
\begin{eqnarray}
Z[\Psi] & = & \int |\Psi|^2,
\end{eqnarray}
where the integral is over all coordinates $\{r_i\}$.  Then, we 
define
\begin{eqnarray}
\gamma & \equiv & \left| {Z[\Psi] - Z[\Psi_{\text{direct}}] \over Z[\Psi]} \right| \\
& = & {\sum_{P\neq I} \int \prod_{i\neq j} |z_i-z_j|^{2m} \phi_{{\bf R}_i}({\bf r}_i)^* \phi_{{\bf R}_i}({\bf r}_{P(i)}) \over \sum_P \int \prod_{i\neq j} |z_i-z_j|^{2m} \phi_{{\bf R}_i}({\bf r}_i)^* \phi_{{\bf R}_i}({\bf r}_{P(i)})}. \label{eq:Rintegral}
\end{eqnarray}
We have used the particle exchange symmetry to reduce the number 
of permutations in each integral from two to one.  Note that when $\gamma$ 
is small, the exchange effect is small and our approximation is good.  

In order to compute $\gamma$ numerically, 
we have developed a 
``permutation 
Monte Carlo method'' \cite{ceperley:boson}, which is essentially the same as the usual 
Monte Carlo simulation method, except for one important difference: 
In a permutation Monte Carlo simulation, not only the electron 
positions ${\bf r}_i$, but also the permutation $P$ is treated 
as a configurational variable which is updated, 
tested, and accepted (or discarded) according to the Metropolis 
algorithm \cite{metropolis}.  Since the integrand in Eq.~(\ref{eq:Rintegral}) 
is a complex quantity, we have separated the phase factor from the 
modulus to sample it.  More specifically, we have averaged 
the phase factor
\begin{equation}
{\phi_{{\bf R}_i}({\bf r}_i)^* \phi_{{\bf R}_i}({\bf r}_{P(i)}) \over \left| \phi_{{\bf R}_i}({\bf r}_i)^* \phi_{{\bf R}_i}({\bf r}_{P(i)}) \right|},
\label{eq:permphase}
\end{equation}
using the rest of the factors in the integrand
\begin{equation}
\prod_{i\neq j} |z_i-z_j|^{2m} \left| \phi_{{\bf R}_i}({\bf r}_i)^* \phi_{{\bf R}_i}({\bf r}_{P(i)}) \right|
\end{equation}
as the statistical weight in the Monte Carlo simulation. 
We note that this permutation Monte Carlo scheme in principle
captures the effects of symmetrization of the wavefunction
exactly.  In practice, however, the precision of the result
depends on how accurately the phase may be sampled.  As described
below, this becomes problematic as $\nu \rightarrow 1/m$. 

\begin{figure}[tb]
\epsfxsize=3.2in
\centerline{ \epsffile{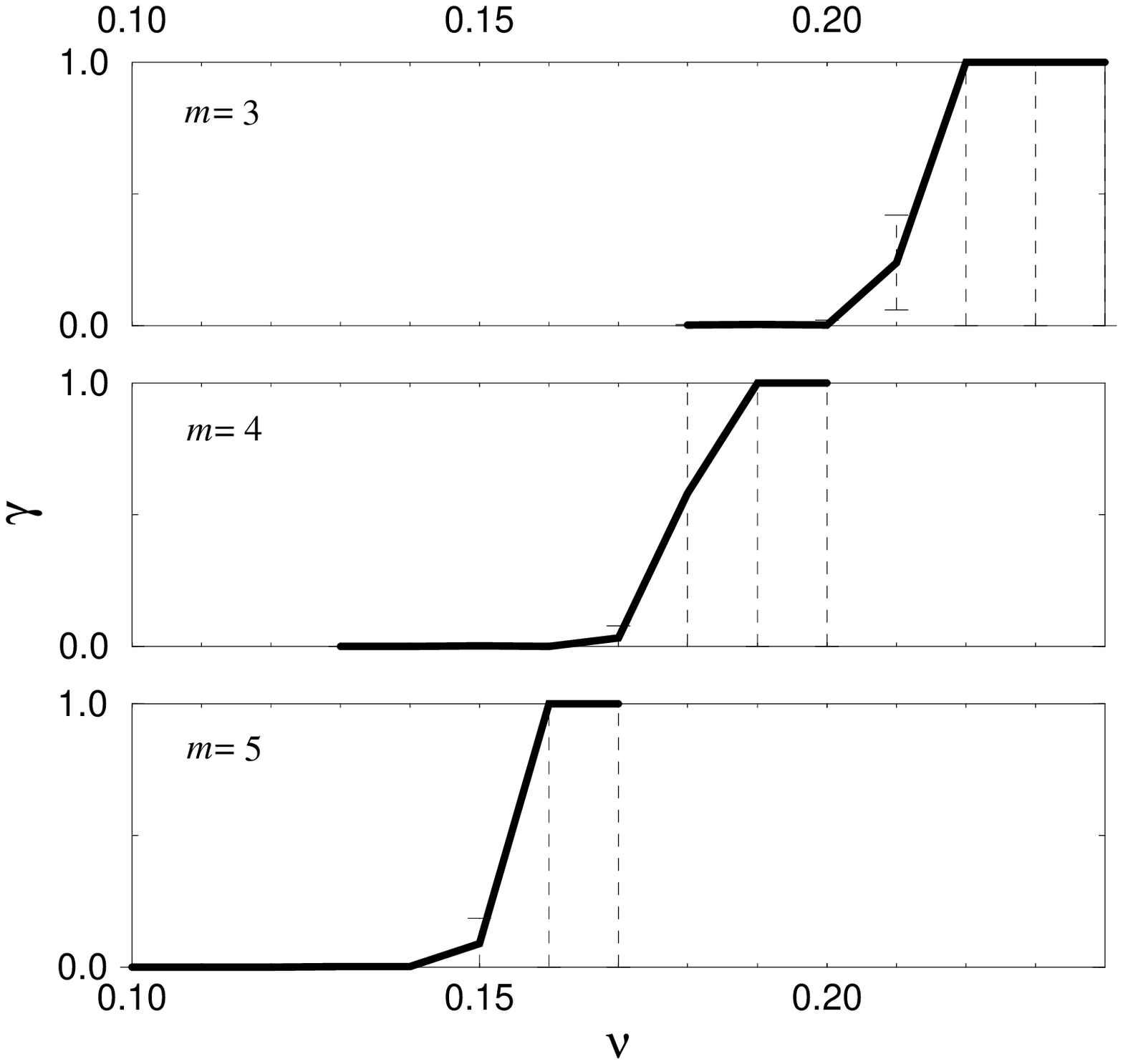} }
\caption{
Relative contribution of the exchange terms in 
the partition function.  $\gamma$ defined in 
(\protect\ref{eq:Rintegral}) 
is plotted for several values of $m$.  
The dashed lines denote errorbars.
}
\label{fig:exchange}
\end{figure}

The results of the permutation Monte Carlo simulation is shown 
in Fig.~\ref{fig:exchange}.  It is clear that $\gamma$ is negligible 
if $\nu$ is less than 0.2, 0.16, and 0.14, for $m=3$, 4, and 5, 
respectively.  Beyond those values of $\nu$, the exchange effect 
can be significant.  The errorbars become quite large as the 
exchange effect gets more and more important.  This is 
because except for the direct term, the phase factor in 
Eq.~(\ref{eq:permphase}) fluctuates very much, while its average 
almost vanishes.  

Note that we have found earlier from Fig.~\ref{fig:energy} 
that the transition between the $m=4$ and the $m=3$ states 
occurs near $\nu=0.16$.  Since the exchange 
effect is negligible up to $\nu=0.16$ when $m=4$, the $m=4$ 
state is guaranteed to have a lower energy than the $m=3$ 
state if $\nu\leq 0.16$, even if the exchange energy is included.  
Above $\nu=0.16$, 
however, it is currently not known whether the exchange effect will 
raise or lower the total energy of the $m=4$ state.  If it 
lowers the energy, there is a possibility that the transition 
from $m=4$ to $m=3$ actually occur at a higher filling 
factor than is shown in Fig.~\ref{fig:energy}.  For $m=5$, 
the exchange effect is negligible up to $\nu=0.14$, which is well 
above $\nu=0.125$ where the transition to the $m=4$ state occurs.  
Therefore, for this transition, our estimate of the transition 
filling factor is accurate.  In principle, 
however, it is possible that a {\em reentrant} $m=5$ phase occurs 
within the $m=4$ ground state if the exchange effect brings the 
energy lower than that of the $m=4$ state above $\nu=0.14$.  
For $m=3$, the exchange effect becomes important well before 
the energy level crosses with that of the $m=2$ state.  However, 
when $\nu\gtrsim 0.2$, not the WC, but the FQH state is the 
ground state, and the energy level crossing between different 
WC states is not physically relevant at $T=0$.  
Finally, we note that for $m=3$, the exchange terms are completely
negligible when $\nu\leq 1/5$.  Therefore, the comparison of the energy 
between our correlated WC state, the Laughlin state, and the 
Lam-Girvin state is valid at $\nu=1/5$ as well as at $1/7$.

\section{Summary}
\label{sec:summary}

In this paper, we have studied the correlated WC in a strong 
magnetic field, which is represented by the product wavefunction 
of the Laughlin-Jastrow factor and the Hartree-Fock wavefunction 
in a triangular lattice.  We have shown that 
extra zeros (ghosts) in the single particle wavefunction 
are necessary to balance the expanding effect of the 
Laughlin-Jastrow correlation.  

The energy of the wavefunction has been calculated using 
Monte Carlo simulations and the Ewald sum method.  Compared to 
other WC trial wavefunctions in the 
lowest Landau level, particularly Lam-Girvin wavefunction, 
the new state is found to be the lowest in energy.  
The Laughlin FQH state has lower energy than our 
wavefunction at $\nu=1/5$ and above, consistent with 
experiment in that the phase transition occurs 
between $\nu=1/7$ and $1/5$.  The energies of the two states 
are, however, very close to each other near $\nu=1/5$.  
This is also consistent with the experimental observation 
of the reentrant insulating phase slightly above $\nu=1/5$, 
considering the cusp in the FQH state energy 
due to quasiparticles and quasiholes.  

Treating $m$ as a variational parameter, we can find the 
value of $m$ which gives our wavefunction the lowest energy.  
Since $m$ takes only discrete integer values, it is found 
that there is a series of first order phase 
transitions between different $m$ states as $\nu$ changes.  
For a given $m$, the spatial fluctuations of the electrons 
grows as $\nu$ increases above $\sim 1/(2m-1)$, eventually 
causing the energy to curve up.  Formation of exchange rings 
is also observed in this limit, which is reminiscent of what occurs
near melting in classical one component plasmas.  

We have also developed a permutation Monte Carlo method in order 
to estimate when the exchange effect becomes important.  Our 
simulation shows that they are negligible up 
to $\nu=0.2$, 0.16, and 0.14 for $m=3$, 4, and 5, respectively.  
This ensures validity of our comparison of the energy 
between different trial wavefunctions at $\nu=1/5$ and 1/7.  
It also shows that when the exchange energy is included, the 
filling factor at which the transition between the $m=4$ and 
the $m=3$ states occurs may be higher, but not lower than our 
estimate in this paper.  However, for the transition from 
the $m=5$ to the $m=4$ states, our estimate of the transition 
filling factor appears to be accurate.

\section*{Acknowledgments}
We wish to acknowledge J.J.\ Palacios, M.C.\ Cha, 
M.R.\ Geller, and I.\ Mihalek for helpful discussions.  
This work was supported by NSF grant DMR 95-03814.

\appendix

\section{Thermodynamic Limit}

In this appendix, we will explain how one can obtain the thermodynamic 
limit from the finite size simulation of the Coulomb energy.  We first 
split the domain of the ${\bf r}'$ integral in Eq.~(\ref{eq:ec}) into 
two regions
\begin{equation}
\int d{\bf r}' = \int_S d{\bf r}' + \int_U d{\bf r}'.
\end{equation}
The first term is an integral over many, but a finite number 
of primitive cells near the center of the disk, 
which we call the ``sampled primitive cells''.  
In this region, the Coulomb energy may be computed directly from 
the Monte Carlo simulations.  The second term concerns the rest 
of the whole 2D plane, the ``unsampled primitive cells'', for 
which we can {\em not} obtain the Coulomb energy directly from 
the simulation.  In order to deal with the second region, 
we use the following trick.  

The contribution from the second region may be written
\begin{equation}
{E_{\text{unsmp}}\over N} = {e^2\over 2} \int_C d{\bf r} \int_U d{\bf r}' {\left< \sum_{i\neq j} \delta({\bf r}-{\bf r}_i)\delta({\bf r}'-{\bf r}_j) \right> \over|{\bf r}-{\bf r}'|},
\label{eq:eunsamp}
\end{equation}
where $\int_C$ means integral over the central primitive cell 
as in Eq.~(\ref{eq:ec}).  
Now, we assume that the correlation is negligible if the 
separation between ${\bf r}$ and ${\bf r}'$ is greater than 
some ``correlation length'' $\xi_c$.  If the distance between 
the unsampled primitive cells ($U$) and 
the central primitive cell ($C$) is greater than $\xi_c$, 
we may use a Hartree approximation and write
\begin{equation}
{E_{\text{unsmp}}\over N} = {e^2\over 2} \int_C d{\bf r} \int_U d{\bf r}' {\rho({\bf r})\rho({\bf r}') \over |{\bf r}-{\bf r}'|},
\label{eq:eunsamprho}
\end{equation}
where we have defined the expectation value of the local density
\begin{equation}
\rho({\bf r}) \equiv \sum_i \langle \delta({\bf r}-{\bf r}_i) \rangle.
\end{equation}
Note that the condition $i\neq j$ in Eq.~(\ref{eq:eunsamp}) is not needed, 
because the domains of ${\bf r}$ and ${\bf r}'$ are exclusive 
so that $\delta({\bf r}-{\bf r}_i)\delta({\bf r}'-{\bf r}_j)\delta_{ij} = 0$.  
Once the local density profile $\rho({\bf r})$ is obtained in the 
central primitive cell from the Monte Carlo simulations, the triangular 
periodicity leads to the density profile 
in the whole 2D plane, and Eq.~(\ref{eq:eunsamprho}) can be explicitly 
computed.  

Although the integral in Eq.~(\ref{eq:eunsamprho}) diverges, 
this divergence is unphysical and 
is easily resolved by recalling that there is neutralizing 
background charge in real samples.  Assuming uniform distribution 
for the positive background charge, the final 
form of the unsampled part is given by
\begin{equation}
{E_{\text{unsmp}}\over N} = {e^2\over 2} \int_C d{\bf r} \int_U d{\bf r}' {[\rho({\bf r})-\overline{\rho}][\rho({\bf r}')-\overline{\rho}] \over |{\bf r}-{\bf r}'|},
\end{equation}
where $\overline{\rho}$ is the average density of electrons.  
The above integral is now well defined and can be computed using 
the Ewald sum method as in Refs.~\ref{ref:bonsall77} and \ref{ref:ewald}.

\begin{thebibliography}{10}
\bibitem{unpub2}
\label{ref:unpub2}
Our initial calculations of interstitial states using quantum Monte Carlo
  calculations reveal that, as in Ref. \onlinecite{zheng94}, Laughlin-Jastrow
  correlations between the interstitial electron and the lattice electrons
  significantly reduce their energy compared to Hartree-Fock estimates of the
  interstitial energy. However, it is found that the energy lowering is not
  great enough to account for what is seen in experiment when correlations
  exist only between the interstitial and lattice electrons, and not among the
  lattice electrons themselves. The instability with respect to interstitial
  formation just below $\nu=1/m$ reported in Ref.~\ref{ref:zheng94} does not
  appear to occur in the more exact Monte Carlo evaluations of the energy.
  Nevertheless, the most significant conclusions of that work -- the lowering
  of interstitial energies by Laughlin-Jastrow correlations and the resulting
  Hall insulating behavior -- are unaffected by this problem.
\end{thebibliography}
\end{document}